\documentclass{iau}
\usepackage{graphicx}
\usepackage{natbib}

\title[Theory of G2 cloud emission] %% give here short title %%
{Theory of G2 cloud \\multi-wavelength emission}

\author[Roman V. Shcherbakov]   %% give here short author list %%
{Roman V. Shcherbakov$^1$}
\affiliation{$^1$Department of Astronomy, University of Maryland, \\College Park, MD 20742-2421, USA\\ email: {\tt roman@astro.umd.edu} }
\pubyear{2013}
\volume{303}  %% insert here IAU Symposium No.
\pagerange{001--004}
\jname{The GC: Feeding and Feedback in a Normal Galactic Nucleus}
\editors{Editors}
\begin{document}

\maketitle

\begin{abstract}
An object called G2 was recently discovered moving towards the supermassive black hole in the Galactic Center.
G2 emits infrared (IR) lines and continuum, which allows constraining its properties.
The question is still unresolved whether G2 has a central windy star or it is a coreless cloud.
Assuming the object is a cloud originating near the apocenter I perform line/continuum IR diagnostics, revisit estimates of non-thermal emission from pericenter passage,
and speculate about future observational prospects. This work is partially reported in \citet{Shcherbakov:2013cl} and partially consists of new ideas discussed at the conference.
%\keywords{Keyword1, keyword2, keyword3, etc.}
%% add here a maximum of 10 keywords, to be taken form the file <Keywords.txt>
\end{abstract}

\firstsection % if your document starts with a section,
              % remove some space above using this command.
%\cite[Anders \& Zinner (1993)]{AndersZinner93}
%(Fig.\,\ref{fig1}).

%\begin{figure}[b]
%% \vspace*{-2.0 cm}
%\begin{center}
% \includegraphics[width=3.4in]{Path.eps}
%% \vspace*{-1.0 cm}
% \caption{Path of pre-solar grains from their stellar sources to the laboratory.}
%   \label{fig1}
%\end{center}
%\end{figure}

%{\underline{\it The lifecycle of pre-solar grains (and maybe interstellar grains in general)}}. Interstellar grains

\section{Origin and Evolution of G2}
The mysterious objects coined G2 was recently discovered in the infrared with Very Large Telescope (VLT) \citep{Gillessen:2012jq}).
The object appears to be moving towards the supermassive black hole (BH) Sgr A* located in the center of the Galaxy.
Emission in Brackett-$\gamma$ (Br$\gamma$), Paschen-$\alpha$ (Pa$\alpha$), and Helium-$I$ (He$I$) lines was distinctly observed as well as the emission in $L'$ and $M$ broad bands.
A detection/non-detection in $Ks$ band with Keck is currently disputed \citep{Phifer:2013ap,Eckart:2013nm}.
A key feature of the source is constant line luminosities. For example, $L({\rm Br}\gamma)$ stayed constant to within $30\%$ from 2004 till 2012.

At present the scientific community is split between two competing hypothesis about what G2 is.
The object was first proposed to be a cloud of gas and dust ionized and irradiated by intense surrounding optical/UV continuum \citep{Gillessen:2012jq}.
Tenuous gas produces recombination lines, while dust emits broadband IR.
The second hypothesis involves a central point source, which produces broadband IR emission \citep{Eckart:2013nm}.
The cold gas component, whose presence is unavoidable, is then either collisionally ionized \citep{Scoville:2013ql} or photoionized.
Cold gas may be produced by collisions and runaway cooling of stellar winds \citep{Cuadrawinds:2008}, photoevaporation of a protoplanetary disk \citep{Murray-Clay:2012zb}
photoevaporation of a disk following encounter of a stellar mass BH with a star \citep{Miralda-Escude:2012al}, a nova outburst \citep{Meyer:2012zf},
and stellar wind from a single young star \citep{Scoville:2013ql,Ballone:2013bv}. Smooth particle hydrodynamics
 simulations of colliding stellar winds presented by Jorge Cuadra show the formation of gas clumps of the right mass.

\section{Infrared Line and Broadband Emission}
Since many estimates of cloud emission are based on photoionization and irradiation by stellar light, it is crucial to compute how much light the bright stars in the Galactic Center produce,
what the spectrum of that light is, and how much the irradiating flux varies along the G2 orbit. I have compiled the known positions \citet{Paumard2006,LuGhez:2009,Gillessen:2009s2},
luminosities, and temperatures \citet{Martins:2007,Martinss2:2008,Cuadrawinds:2008} of the most massive stars in the region
and computed the expected starlight spectra/fluxes at the G2 positions. IRS16NW, IRS16C, and IRS16SW stars provide most of the flux, while situated at $1-4$~arcsec distance from Sgr A*.
Then, as the G2 distance to Sgr A* decreases from $0.6$~arcsec in 2004 to $0.3$~arcsec in 2011, the irradiating flux from these stars changes very little.
A $30\%$ increase of the irradiating flux between 2004 and 2011 is largely associated with a growing contribution of S0-2 star on the orbit with a semi-major axis $0.12$~arcsec \citep{Gillessen:2009s2}.
The calculations were performed with $20,000$~K temperature of IRS16 stars \citep{Martins:2007} and $30,000$~K temperature of S0-2 star \citep{Martinss2:2008}.
Total irradiating flux at G2 position in 2011 is about $F=4.0\times10^{4}{\rm erg~s}^{-1}{\rm cm}^{-2}$.
Photoionization calculations were self-consistently conducted with CLOUDY code \citep{Ferland:2013cp}.

As shown by \citet{Eckart:2013nm}, emission from a single central young star can explain broadband IR observations of G2.
However, even a small addition of dust to cold gas with sub-Galactic dust-to-gas ratio can readily reproduce $L'$ and $M$ luminosities \citep{Shcherbakov:2013cl}, while also being consistent
with $Ks$ band magnitude. Small dust grains with size $\sim10$nm are self-consistently heated to $500$~K. Their emission naturally explains the observed $L'-M$ color.
Heavier dust grains radiate disproportionately in $M$ band, thus providing worse fit to $L'-M$ color.
However, "ISM" size distribution of dust grains implemented in CLOUDY provides acceptable agreement with broadband G2 spectrum.
Gas line emission is trickier to model.

I have discovered that the cold gas cloud could either be optically thin or optically thick to the ionizing continuum.
In the optically thick case the density and mass of the cloud are larger.
Line luminosity in this case is proportional to the absorbed power
\begin{equation}
L({\rm Br}\gamma)_{\rm thick}\propto F S,
\end{equation} where $S$ is the surface area of the cloud.
Then, as the irradiating flux $F$ changes very little with time, a constant or slightly decreasing surface area leads to a constant $L({\rm Br}\gamma)$ consistently with observations.
In the optically thin case the entire cloud is heated up to $T\approx10^4$~K and emits recombination lines.
The correspondent luminosity is
\begin{equation}
L({\rm Br}\gamma)_{\rm thin}\propto n m_{\rm cloud}T^{-1},
\end{equation} where $n$ is the cloud density. Then constant cloud density ensures constant $L({\rm Br}\gamma)$.

The boundary between the optically thin and optically thick models and the evolution of line luminosity with time depend on the cloud shape.
I considered three different cloud shapes assuming the cloud preserves its mass: spherical, tidally distorted, and magnetically arrested.
Each model starts as a spherical cloud near the apocenter at $r=1$~arcsec three-dimensional distance from Sgr A* \citep{Gillessen:2013pe}.
The spherical cloud has constant shape and density, thus automatically producing constant $L({\rm Br}\gamma)$.
However, the self-gravity of G2 is weak and the cloud is not expected to preserve its shape.
A tidally distorted cloud is elongated in the radial direction as $L\propto r^{-1/2}$, where $r$ is the distance to Sgr A*.
The perpendicular size shrinks as $\rho\propto r$, thus density sharply rises as G2 approaches the pericenter.
The optically thin tidally distorted cloud cannot preserve constant line luminosity and is then disfavored by observations.
Similarly large changes in line luminosity are found in numerical simulations of stellar wind from a single central star scenario \citep{Ballone:2013bv}.
The magnetically arrested cloud amplifies radial magnetic field with help of magnetic flux conservation. In this model the magnetic force balances the projected gravitational force and
the perpendicular size varies as $\rho\propto r^{5/8}$. Density increases only weakly with time and the changes of $L({\rm Br}\gamma)$ are consistent with data
for the optically thin magnetically arrested cloud. The model, which best fits IR data, has the initial cloud radius $R_{\rm init}\approx43$~mas and density $n_{\rm init}\approx4\times10^4{\rm cm}^{-3}$.

It was recently claimed \citep{Burkert:2012ca} that the cloud would be too extended in projection, if it started at the apocenter.
However, first, starting at $r=1$~arcsec leads to the line velocity spread consistent with observations \citep{Gillessen:2012jq}.
Secondly, revised orbital parameters of the G2 cloud presented by Leo Meyer at the conference make G2 move at a smaller angle to the line of sight,
effectively hiding the radial extent of the cloud.  An additional concern is that the Kelvin-Helmholtz (KH)and Rayleigh–Taylor (RT) instabilities as well as the shock from the ambient medium would
destroy the cloud moving from the apocenter \citep{Gillessen:2012jq,Burkert:2012ca}.  However, even weak initial magnetic field helps to save the cloud.
The object becomes stable against the ambient medium shock and the timescales of RT/KH instabilities become large.

\section{Radiation at the Pericenter}
%\todo{Done till here}
Two radiative effects were predicted to take place, when the G2 cloud passes through the pericenter.
First, the bow shock into the ambient gas is expected to accelerate particles, which should emit synchrotron radiation with observable radio power \citep{Narayan:2012jl,Sadowski:2013mn}.
The expected radio flux is $\sim10$ times larger than the quiescent Sgr A* flux at $1.4$~GHz, but no brightening was observed at all \citep{Bower:2013za}.
Such discrepancy can be resolved within the compact scenario of stellar wind launched by a single star \citep{Ballone:2013bv}.

However, more realistic flux estimates show that my gas cloud model readily leads to radio non-detection.
The cross-section of my best-fitting magnetically arrested cloud at the pericenter is about $5$ times lower than the fiducial cross-section used by \citet{Narayan:2012jl}.
The density profile of the ambient hot gas with a slope $\beta=0.85$ \citep{Shcherbakov:2012appl} is shallower than the widely used profile $n\propto r^{-1}$,
so that the ambient gas density is a factor of $2$ lower at the pericenter. The shock initiated but the G2 cloud is not perpendicular, but is very oblique.
The particle acceleration efficiency is largely unknown for the oblique shocks and the fiducial efficiency of $5\%$ assumed by \citet{Narayan:2012jl} might not be achieved.
In sum, the peak of G2 cloud radio flux is only expected to match Sgr A* flux at $1.4$~GHz for the fiducial efficiency of $5\%$, but lower efficiency values readily lead to non-detection.

The second radiative effect is the X-ray emission from the cloud material heated by the shock from the ambient gas.
The expected X-ray power is above the detection threshold \citep{Gillessen:2012jq}, but this effect only works, if G2 is unmagnetized.
Even a weakly magnetized cloud becomes stable against the shock from the ambient gas, which leads to no X-ray brightening.
Similarly, the tidal shock, when parts of G2 on slightly different trajectories slam into each other, is expected to be stalled by the magnetic field.
Then the most promising heating mechanism, when G2 approaches the pericenter, is dissipation of the accumulated B-field.
Such dissipation leads to gas and dust heating and the increase of IR broadband emission.
Correspondingly, Br$\gamma$ luminosity would increase in the optically thick models, but would stay about constant in the optically thin models.

\section{What to Expect after Pericenter}
The computed best-fitting cloud model has a mass $m_{\rm cloud}\approx13M_{\odot}$, which is about $4$ times heavier than the fiducial cloud in \citet{Gillessen:2012jq}.
Simultaneously, the cloud has a smaller perpendicular cross-section and substantial magnetization. The ambient gas density may have been previously overestimated by a factor of $2$.
All four factors work to reduce the efficiency of cloud interaction with the ambient gas.
Then, instead of being quickly slowed down, destroyed by shear, and evaporated by conduction \citep{Anninos:2012kz,Burkert:2012ca,Gillessen:2012jq},
the cloud evolves ballistically. The ballistic evolution implies a long timescale of debris fallback onto the central object.
Then, the brighter state of BH accretion predicted due the infall of debris \citep{Moscibrodzka:2012ax} may be postponed.
Sgr A* might not flare up substantially even though the cloud is heavier, as the return of the debris is spread over the longer time.

\acknowledgements
RVS is supported by NASA Hubble Fellowship grant HST-HF-51298.01.

%\bibliographystyle{plain}
%\bibliography{refs}

\end{document}